\documentclass[10pt]{IEEEtran}
\usepackage{pgfplots}
\DeclareUnicodeCharacter{2212}{−}
\usepgfplotslibrary{groupplots,dateplot}
\usetikzlibrary{patterns,shapes.arrows}
\pgfplotsset{compat=newest}
\usepackage{cite}
\usepackage{amsmath,amssymb,amsfonts}
\usepackage{algorithmic}
\usepackage{graphicx}
\usepackage{textcomp}
\usepackage{xcolor}
\usepackage{cite}
\usepackage{amsmath,amssymb,amsfonts}
\usepackage{algorithmic}
\usepackage{graphicx}
\usepackage{textcomp}
\usepackage{xcolor}
\usepackage{cite}
\usepackage{epsfig}
\usepackage{fancyhdr}
\usepackage{color}
\usepackage{graphicx}
\usepackage[linesnumbered,ruled,vlined]{algorithm2e}
\usepackage{amsmath}
\usepackage{algorithmic}
\usepackage[T1]{fontenc}
\usepackage{times}
\usepackage{epsfig}
\usepackage{fancyhdr}
\usepackage{framed}
\usepackage{multirow}
\usepackage{url}
\usepackage{balance}
\usepackage{booktabs}
 \usepackage{graphicx}
\usepackage{listings}
\usepackage{caption}
\usepackage{upquote}
\usepackage{xcolor}
\usepackage{comment}
\usepackage{subfigure}
\usepackage{mathtools}
\usepackage{ragged2e}
\usepackage{lastpage}
\usepackage{academicons}

\usepackage{hyperref}
\usepackage{array}
\usepackage{epsfig}
\usepackage{fancyhdr}
\usepackage{color}
\usepackage[linesnumbered,ruled,vlined]{algorithm2e}
\usepackage{amsmath}
\usepackage{times}
\usepackage{framed}
\usepackage{multirow}
\usepackage{url}
\usepackage{balance}
\usepackage{booktabs}
\usepackage{listings}
\usepackage{caption}
\usepackage{upquote}
\usepackage{comment}
\usepackage{subfigure}
\usepackage{mathtools}
\usepackage{ragged2e}
\usepackage{lastpage}
\usepackage{academicons}
\usepackage{adjustbox}
\usepackage{hyperref}
\usepackage{adjustbox}
\usepackage{academicons}
\usepackage{hyperref}
\usepackage{tabularx}
\newcommand{\nik}[1]{\textcolor{blue}{\textbf{#1}}}
\newcommand*\CircledText[1]{\tikz[baseline=(char.base)]{
            \node[shape=circle,draw,inner sep=0.5pt] (char) {#1};}}
\renewcommand{\thefootnote}{\textit{\alph{footnote}}}

\ifCLASSINFOpdf
\else
\fi

\newcommand{\unnumberedfootnote}[1]{
    \renewcommand{\thefootnote}{}
    \footnotetext{#1}
    \renewcommand{\thefootnote}{\arabic{footnote}}
}
\begin{document}
\newcommand{\papername}{MUSIC-lite}
\newcommand{\todo}[1]{\textcolor{red}{#1}}

\title{\papername{}: \texorpdfstring{Efficient MUSIC using Approximate Computing: An OFDM Radar Case Study}}

\author{
  \IEEEauthorblockN{Rajat Bhattacharjya\IEEEauthorrefmark{1}, Arnab Sarkar\IEEEauthorrefmark{2}, Biswadip Maity\IEEEauthorrefmark{1}, and Nikil Dutt\IEEEauthorrefmark{1}
   }\\
   \IEEEauthorblockA{\IEEEauthorrefmark{1}Department of Computer Science; University of California, Irvine}\\
  \IEEEauthorblockA { \IEEEauthorrefmark{2}Department of Electronics and Electrical Communication Engineering; Indian Institute of Technology, Kharagpur}\\
  \IEEEauthorrefmark{1}\{rajatb1, maityb, dutt\}@uci.edu, \IEEEauthorrefmark{2}arnabsarkar@kgpian.iitkgp.ac.in
  }

%\newcommand{\short}{ACLA }
%\newcommand{\shortx}{ACLA}



%\author{\IEEEauthorblockN{Rajat Bhattacharjya\IEEEauthorrefmark{1},
%Telephone: (800) 555--1212, Fax: (888) 555--1212}


\maketitle
%\IEEEoverridecommandlockouts
%\IEEEpubid{\makebox[\columnwidth]
%{979-8-3503-8122-1/24/\$31.00~\copyright2024 IEEE \hfill}
%\hspace{\columnsep}\makebox[\columnwidth]{ }}
%\IEEEpubidadjcol
\begin{abstract}
Multiple Signal Classification (MUSIC) is a widely used Direction of Arrival (DoA)/Angle of Arrival (AoA) estimation algorithm applied to various application domains such as autonomous driving, medical imaging, and astronomy. However, MUSIC is computationally expensive and challenging to implement in low-power hardware, requiring exploration of trade-offs between accuracy, cost, and power. We present \papername{}, which exploits approximate computing to generate a design space exploring accuracy-area-power trade-offs. This is specifically applied to the computationally intensive singular value decomposition (SVD) component of the MUSIC algorithm in an orthogonal frequency-division multiplexing (OFDM) radar use case. \papername{} incorporates approximate adders into the iterative CORDIC algorithm that is used for hardware implementation of MUSIC, generating interesting accuracy-area-power trade-offs. Our experiments demonstrate \papername{}'s ability to save an average of 17.25\% on-chip area and 19.4\% power with a minimal 0.14\% error for efficient MUSIC implementations.

\unnumberedfootnote{Paper accepted at ESWEEK-CASES 2024. Authors' version of the work posted for personal use and not for redistribution. The definitive version will be available in IEEE Embedded Systems Letters.}
\begin{comment}

Multiple Signal Classification (MUSIC) is a widely used Direction of Arrival (DoA)/ Angle of Arrival (AoA) estimation algorithm applied to various application domains such as autonomous driving, medical imaging, and astronomy. 
However, MUSIC is computationally expensive and challenging to implement in low-power hardware, requiring exploration of tradeoffs between accuracy, cost and power. 
We present \papername{} 
that exploits approximate computing to generate a design space exploring accuracy-area-power tradeoffs for  
the computationally intensive singular value decomposition (SVD) component of MUSIC.
%to address this challenge by introducing approximate adders inside the computationally intensive block of MUSIC, i.e, the singular value decomposition (SVD). 
\papername{} incorporates approximate adders into the iterative CORDIC algorithm  used for hardware implementation of  MUSIC,
\nik{generating interesting accuracy-area-power tradeoffs. 
Our experiments on CORDIC implementations of SVD demonstrate
 \papername{}'s ability to  save an average of 17.25\% on-chip area and 19.4\% power
with a minimal 0.14\%  error for efficient MUSIC implementations.} 
%on hardware, and thereby achieve hardware efficiency at a marginal loss of accuracy. 
\end{comment}
\end{abstract}



% \fi
\IEEEpeerreviewmaketitle

\section{Introduction}

\label{sec:intro}
%Context: what is the system/application, why do I care?
%Challenge: what’s so hard? What’s the problem to solve? 
%Opportunity- gap in existing work?
%Contribution: how are you going to fill the gap in solving the problem? What do you bring that’s novel?

%Context: what is the system/application, why do I care?
Multiple Signal Classification (MUSIC) is a widely used signal processing technique applied to Direction of Arrival (DoA) or Angle of Arrival (AoA) estimation problems~\cite{music}. 
MUSIC finds multiple applications in diverse fields such as 
localization in autonomous vehicles~\cite{use3}, signal analysis in biomedical sensors~\cite{biomed, use2}, seismic monitoring systems~\cite{seismic}, and joint radar-communication systems ~\cite{musicuse}.  

%Challenge: what’s so hard? What’s the problem to solve? 
MUSIC relies on matrix decomposition techniques like eigenvalue decomposition (EVD) and singular value decomposition (SVD) during the signal processing stage.
Traditionally, these matrix decomposition techniques are computationally expensive and power-hungry~\cite{pak,admm} thereby negatively impacting the deployment of MUSIC on resource-constrained embedded platforms.
%Traditionally, matrix decomposition techniques involves matrix multiplications, making MUSIC computationally expensive and power hungry~\cite{pak, admm}.
%Additionally, these signal matrices used are often sparse, complicating the design of power-efficient hardware implementations.  
%Opportunity- gap in existing work?
Therefore, we need solutions that can implement MUSIC efficiently on low-power hardware. 
Existing works focus on making MUSIC implementation efficient by reducing the number of snapshots~\cite{snap}, or using a less-precise covariance matrix~\cite{onebit}. 
%However, these approaches have limited use cases (e.g., Uniform Linear Antenna Array), and prioritize reducing latency over power-efficiency.
However, these approaches suffer from high error rates.
%\todo{Rajat: Please add one line about where they lack}

%Contribution: how are you going to fill the gap in solving the problem? What do you bring that’s novel?
In this paper, we present  \papername{} that exploits approximate computing~\cite{surveyapprox,cesa,locate} to reduce the computational complexity of MUSIC, while generating power efficient solutions with minimal loss of accuracy. 
We apply \papername{} to an orthogonal frequency-division multiplexing (OFDM) radar pipeline case study that deploys
%Specifically, we focus on introducing 
approximate arithmetic circuits (adders) to generate a design space trading off power, accuracy and cost (area).
%into an 
%orthogonal frequency-division multiplexing (OFDM) radar pipeline (shown in Figure~\ref{fig:outline}) to evaluate the efficacy of 
 %\papername{}. 
%We propose the use of a hardware-friendly algorithm to implement SVD stage in MUSIC, and explore approximate computing~\cite{surveyapprox} to enhance the power-efficiency of MUSIC.
%Specifically, we introduce approximate arithmetic circuits (adders) into an orthogonal frequency-division multiplexing (OFDM) radar pipeline (shown in Figure~\ref{fig:outline}). 
The OFDM radar pipeline case study highlights several benefits: 1) The versatility of the OFDM waveform -- used in both communication and radar systems -- allows for a unified and comprehensive analysis using  \papername{} 
%. This versatility is instrumental in the development of 
to develop efficient joint radar-communication systems
%, which are utilized 
across multiple application domains, including automotive, healthcare, and military applications;
2) OFDM systems exhibit high spectral efficiency, adaptability to varying bandwidth requirements, and robustness to Doppler shifts, making them invaluable across diverse applications and challenging environments.

We reduce the complexity of MUSIC by leveraging hardware-efficient approximations in Singular Value Decomposition (SVD). Due to the inherent error-resilience of SVD~\cite{svdresil}, introducing approximations can substantially reduce on-chip power consumption and area. To achieve this, we utilize the CORDIC algorithm (CA)~\cite{cordic} for the hardware implementation of SVD, as it primarily relies on iterative shifts and additions. By integrating approximate adders into CA, we further enhance the hardware efficiency of the SVD process.

Our experimental results show that 
%the proposed approach 
\papername{}
saves 17.25\% on-chip area and 19.4\% power with an average minimal error of 0.14\% in the positive SNR region. 
To the best of our knowledge, \papername{} is the first to  explore the use of approximate arithmetic circuits in MUSIC for area and power-efficiency.

\vspace{-3ex}

\begin{figure}
\includegraphics[width=0.48\textwidth]{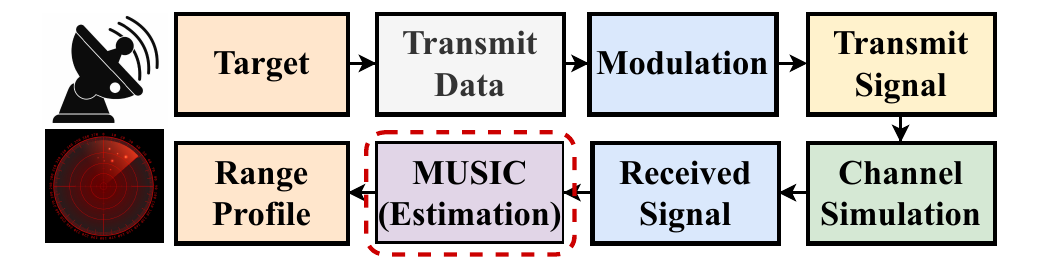}
\caption{Orthogonal frequency-division multiplexing (OFDM) radar processing pipeline. We approximate the computationally intensive MUSIC block highlighted in red.}
\vspace{-3ex}
\label{fig:outline}
\end{figure}

\section{Background}
\label{sec:sys}
%An OFDM-Radar processing pipeline is taken into account as shown in Figure~\ref{fig:overview} with the following properties (Table~\ref{tab:sys}). Multiple Signal Classification (MUSIC) is performed to extract range of multiple targets.

We begin with a brief background on
(1) MUSIC, 
(2) Singular value decomposition (SVD) implementation  using the Golub-Kahan algorithm, and 
(3) Hardware realization of the Golub-Kahan algorithm using CORDIC.
% For a uniform linear array of antennae, MUSIC works by acquiring signals, constructing the covariance matrix, performing singular value decomposition to extract signal and noise subspace and finally doing a peak search by finding minima in the noise subspace. The flow can be seen in Figure~\ref{fig:method}. 

(1) \textbf{MUSIC} -- for a uniform linear antenna array -- acquires signals, constructs the covariance matrix, performs the SVD to extract the signal and noise subspace, and then finally performs a peak search by finding minima in the noise subspace. 
Figure~\ref{fig:method} outlines the MUSIC flow (using CORDIC for SVD).
%is shown in Figure~\ref{fig:method}.
Mathematically, MUSIC spectra is represented by:

\begin{equation}
\label{MUSIC_eq}
P_{MU}(\theta) = \frac{1}{a^{*}(\theta)E_{N}E_{N}^{H}a(\theta)}
\end{equation}

Where $P_{MU}(\theta)$ is the MUSIC spectra, $E_{N}$ is the matrix whose columns are the eigenvectors of the noise subspace, 
and $a(\theta)$ are the elements of the array steering matrix.
Equation~\ref{MUSIC_eq} 
%will lead to 
generates
large peaks observed at the direction/angle of arrival ($\theta$),
as shown in the MUSIC Spectrum of Figure~\ref{fig:method}.

(2) \textbf{Singular Value Decomposition (SVD)} is the most computationally intensive part of MUSIC for calculating the singular values as stated in Section~\ref{sec:intro}. 
Hence, we direct our focus on SVD. A fast and numerically stable method to calculate SVD is the Golub-Kahan algorithm~\cite{ref_golub}. 
The Golub-Kahan algorithm begins by transforming the given matrix \( A \) into a bidiagonal form. 
It then diagonalizes this bidiagonal matrix. 
The resulting diagonal elements are the singular values of \( A \). The transformation of the matrix to its bidiagonal and diagonal form is achieved using Givens rotations. 
%(as rotations are unitary transformations and maintain the orthogonality of U and V matrices).

(3) \textbf{ The CORDIC algorithm (CA)} efficiently realizes  the Golub-Kahan algorithm in hardware by using  shifts and additions for the Givens rotations.
%effectively using shifts and additions. 
%On hardware, Cordic Algorithm (CA) has emerged as a popular technique to implement the givens rotations. This is because CA uses shifts and additions to perform rotations.\\
CA iteratively shifts and updates the $x$ and $y$ coordinates to achieve the required rotation as shown below:
\begin{align*}
x_{i+1} &= x_i - d_i \cdot y_i \cdot 2^{-i} \\
y_{i+1} &= y_i + d_i \cdot x_i \cdot 2^{-i} \\
z_{i+1} &= z_i - d_i \cdot \arctan(2^{-i})
\end{align*}

\begin{itemize}
    \item \( x_{i+1} \): The x-coordinate after the \(i\)-th iteration.
    \item \( y_{i+1} \): The y-coordinate after the \(i\)-th iteration.
    \item \( z_{i+1} \): The angle accumulator after the \(i\)-th iteration.
    \item  \(d_{i}\): Direction factor.
\end{itemize}

Since CA involves iterative additions for convergence, approximating the addition operations enables the exploration of power-efficient design points that can also meet the accuracy requirements of the SVD algorithm.
\section{\papername{} Methodology}
\begin{figure}
\includegraphics[width=0.48\textwidth]{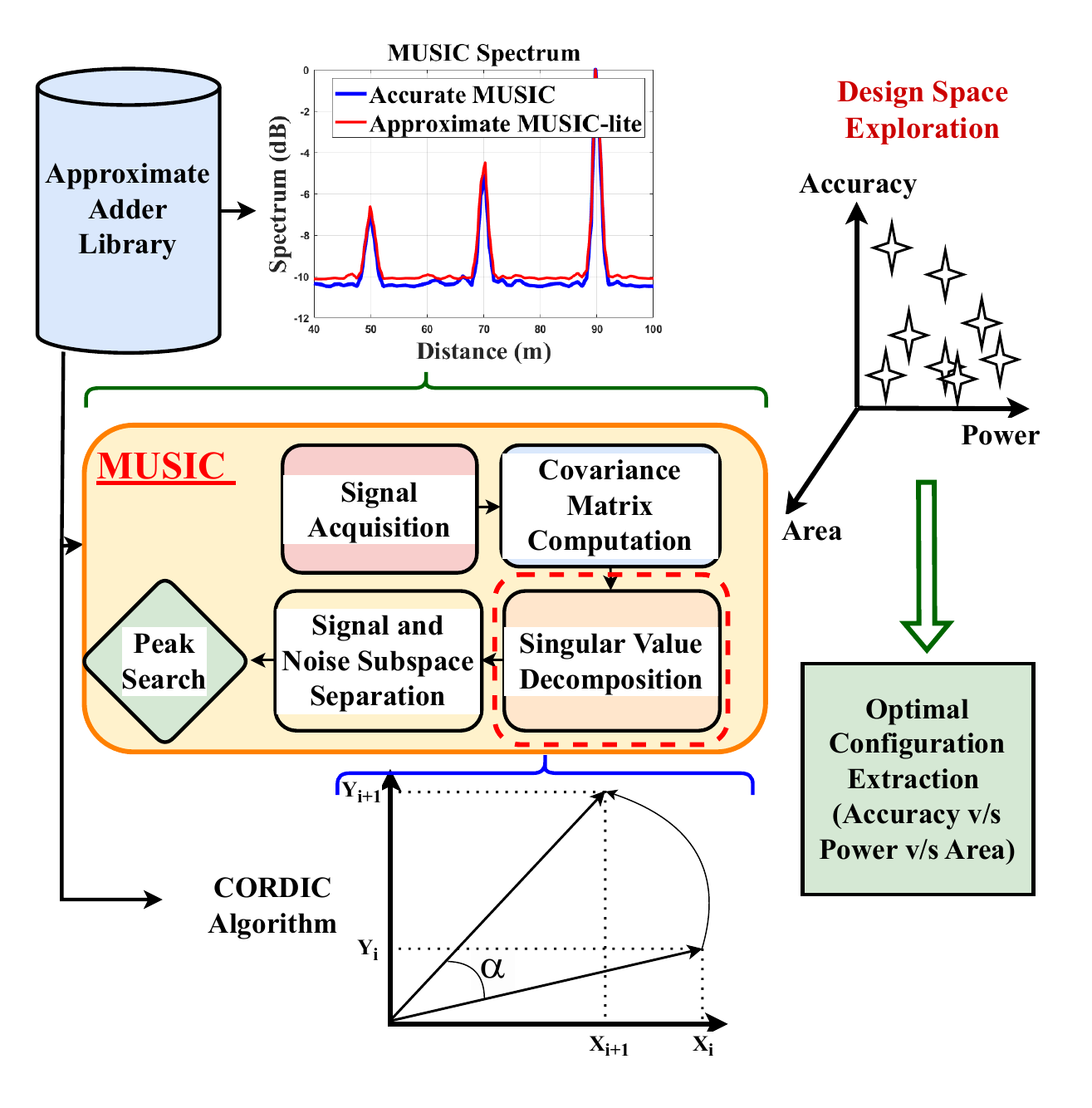}
\caption{Approximation using \papername{} }
\label{fig:method}
\vspace{-3ex}
\end{figure}
\label{sec:sw}
%In this section, we study the OFDM Radar pipeline in consideration.%which MUSIC is approximated. Furthermore, we list the steps that we follow to achieve that. 
We  present our \papername{} methodology (Figure~\ref{fig:method}) for  the OFDM radar case study 
%presented the OFDM radar pipeline in 
(Figure~\ref{fig:outline}),
with the goal of demonstrating the utility of
\papername{} in supporting the design space exploration of
alternative MUSIC implementation across  different accuracy-power-area design points for end-to-end analysis of the OFDM radar pipeline.
%In this section, we illustrate the \papername{} methodology using the OFDM radar case study.
%The OFDM radar  pipeline provides us with the opportunity to evaluate our proposed approach in a relevant end-to-end use case and thereby understand the overall benefits which approximating MUSIC provides. 
Table~\ref{tab:sys} shows the various parameters associated with the OFDM radar pipeline in our study.

% As mentioned in Section \ref{sec:intro}, we focus on MUSIC (highlighted in red in Figure~\ref{fig:outline}), and introduce approximations in SVD~\cite{pak}. 
% In particular, we implement the SVD using CA, and approximate the addition operations inside CA. 

The \papername{} methodology (Figure~\ref{fig:method}) can be broken down into 3 major steps: (A) Functional Validation, (B) Hardware Implementation, and (C) Design Space Exploration.
\subsection{Functional Validation}
\label{sec:fn}
First, at the software level, we introduce behavioral models of approximate adders in the radar pipeline. We use MATLAB to perform our experiments.
We use the EvoApprox Library~\cite{evoapprox} to source the approximate adders. 
We select the approximate adders based on the their individual error metrics %they have 
(e.g., Error Percentage (EP), Worst-Case Error (WCE), Mean Absolute Error (MAE)~\cite{stat}). 
We run the entire pipeline 100 times for an SNR range of -5 to 15 dB with a step-size of 5. 
This exploration enables us to understand the effect of approximations at different noise levels and thereby provides a distinct boundary between conditions supporting approximations and vice-versa. 
%For practical applications, we assume the SNR to be in the positive range to proceed to the next step in our methodology.

%\todo{Rajat:The accuracy metric is..} 
The accuracy metric is the estimated target's range in meters when deploying  various approximate adders for MUSIC in the OFDM radar pipeline. If the approximated output level (target's range) falls within a user-defined  quality window, we then proceed with  hardware implementation. 

%This helps us discard the adders which do not fall under the acceptable quality window.

\subsection{Hardware Implementation}
\label{sec:hi}
We evaluate the improvements in on-chip area and power efficiency by implementing a CORDIC core in hardware using Verilog HDL. 
Once we have the adders that fall within the application's acceptable quality window from the previous step, we incorporate them into the CORDIC core which is responsible for performing the SVD in MUSIC. 
Finally, we synthesize the approximated CORDIC cores using Synopsys Design Compiler with a $45 nm$ technology node.

Next, the design space of the OFDM radar pipeline is explored by varying the approximate adders.

\subsection{Design Space Exploration}
Now that we have both accuracy and hardware statistics from Sections~\ref{sec:fn} and~\ref{sec:hi}, we can identify optimal design points and the trade-offs associated with them. 
To determine the optimality of a design point, we consider user-defined quality constraints for both accuracy and hardware statistics (area, power). Thus, based on a predefined accuracy threshold and area and power budget, we make selections of approximate designs satisfying all three requirements. 

\begin{table}[]

\caption{OFDM system parameters for Fig.~\ref{fig:outline}}
\label{tab:sys}
\begin{adjustbox}{width=\columnwidth,center}
\begin{tabular}{|l|l|}
\hline
\textbf{System Parameter/Block} & \textbf{Value/Property}               \\ \hline
Carrier frequency               & 30 GHz                                \\ \hline
Number of subcarriers           & 32                                    \\ \hline
Number of symbols               & 16                                    \\ \hline
Subcarrier spacing              & 960 kHz                               \\ \hline
Elementary OFDM symbol duration & 1.04 $\mu$s                          \\ \hline
Cyclic prefix duration          & 0.26 $\mu$s                          \\ \hline
Total symbol duration           & 1.3 $\mu$s                           \\ \hline
Modulation                      & 4-QAM                                 \\ \hline
%Zadoff-Chu sequence index       & 5                                     \\ \hline
%Zadoff-Chu sequence root        & 29                                    \\ \hline
Target position                 & 50 m                                  \\ \hline
Target velocity                 & 20 m/s                                \\ \hline
Channel                         & Additive White Gaussian Noise Channel \\ \hline
\end{tabular}
\end{adjustbox}
\vspace{-3ex}
\end{table}
\section{Evaluation}
\label{sec:eval}
We now present the efficacy of \papername{} in the end-to-end OFDM radar case study.
We approximate the addition operations inside the SVD (CORDIC algorithm) of MUSIC and then obtain results for 
the target's range across an end-to-end pipeline as shown in Figure~\ref{fig:outline}.  
The target is situated at a distance of $50 m$. 
We evaluate 15 16-bit sign-extended approximate adders from the EvoApprox Library~\cite{evoapprox} with diverse accuracy metrics~\cite{stat}. 
%Functional Validation provides us with accuracy statistics for the application in consideration, i.e, OFDM-radar pipeline. For hardware implementation, we design our own IFFT core and approximate the adder-multiplier pairs in the butterfly unit so as to obtain area and power results. Finally, we conduct the DSE which helps us explore design points satisfying user-defined quality constraints corresponding to the parameters of accuracy, area, and power, thereby helping us generate low-power and energy-efficient radar processors.
\subsection{Accuracy Analysis}
\label{sec:acc}
\begin{figure}
    \includegraphics[width=0.48\textwidth]{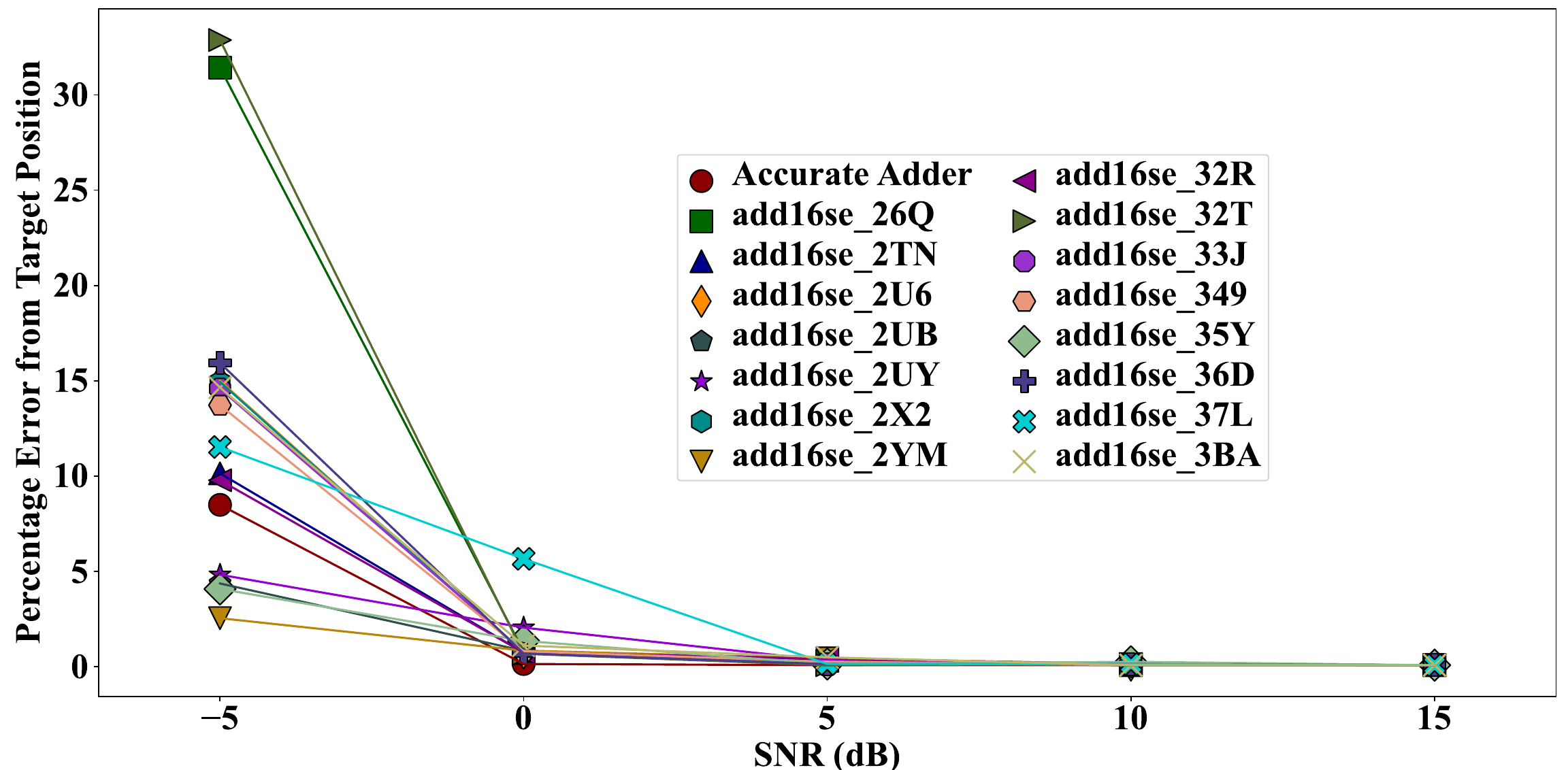}
    \caption{Accuracy statistics using various adders (accurate and approximate). Range averaged for 100 runs per SNR for each adder.}
    \label{fig:acc}
    \vspace{-3ex}
\end{figure}

%To show that approximations have negligible impact on application accuracy, 
Figure~\ref{fig:acc} shows that our approximations have negligible impact on application accuracy.
The figure shows
the percentage error for range estimation (Y-axis) while using both accurate and approximate adders across an SNR range of -5 to +15 dB (X-axis). 
The range values obtained per SNR are averaged across 100 runs per adder. 

We make 4 major observations. 
1) We see that for negative SNR, all adders (including the accurate adder) perform poorly with high error. %, including the accurate adder. 
However, even in this case we see that three approximate adders perform better than the accurate adder, with the best approximate adder (\texttt{add16se\_2YM}) showing only a 2.534\% error from the target's actual position. 
2) We observe that at SNR = 0 dB, the accurate adder performs best with a deviation of 0.13\% from the target's range, and among all approximate adders, \texttt{add16se\_32T} performs best with a percentage error of 0.66\%. 
3) As we move towards higher positive SNR, all adders more or less perform the same (including the accurate adder) in estimating target's range, with the highest percentage error being 0.421\% among all adders. 
4) For higher positive SNR (5 to 15 dB), among all approximate adders, \texttt{add16se\_33J} performs the best with an average error percentage of 0.088\%,  although it has an inherent error probability of 0.25. 
These observations lead to the conclusion that since there are many dynamic interactions happening over an end-to-end system, such a study is critical for deploying approximate adders in evaluating end-to-end application accuracy.
%otherwise we might arrive to a different conclusion based on the listed error metrics for the adders in consideration.
\begin{figure}
\begin{center}
    \includegraphics[width=0.46\textwidth]{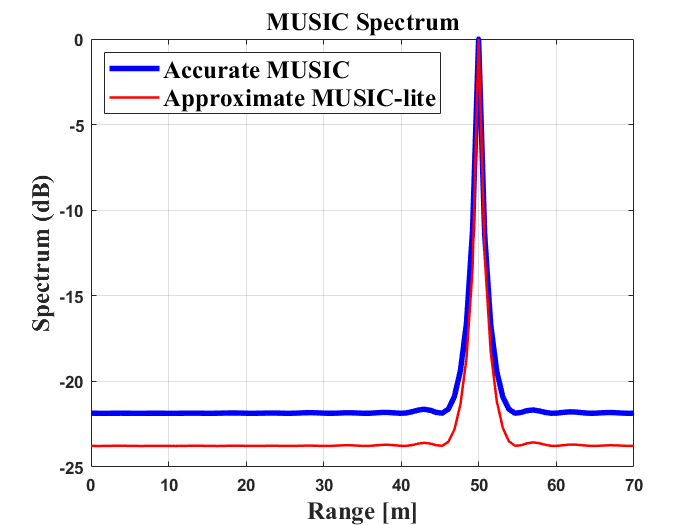}
    \caption{Target's range profile at SNR = 10 dB with accurate adder; and approximate adders \texttt{add16se\_2U6}, \texttt{add16se\_2TN}.}
    \label{fig:music}
    \end{center}
    \vspace{-4ex}
\end{figure}
Overall, Figure~\ref{fig:music} shows that approximate adders have negligible impact on application accuracy, obtaining similar statistics as with using accurate adders,
%(shown in Figure~\ref{fig:music}), 
with an error percentage of 0.14\% (from the target's position) for practical radar scenarios that have positive SNR.

\begin{figure}[b]
    \includegraphics[width=0.48\textwidth]{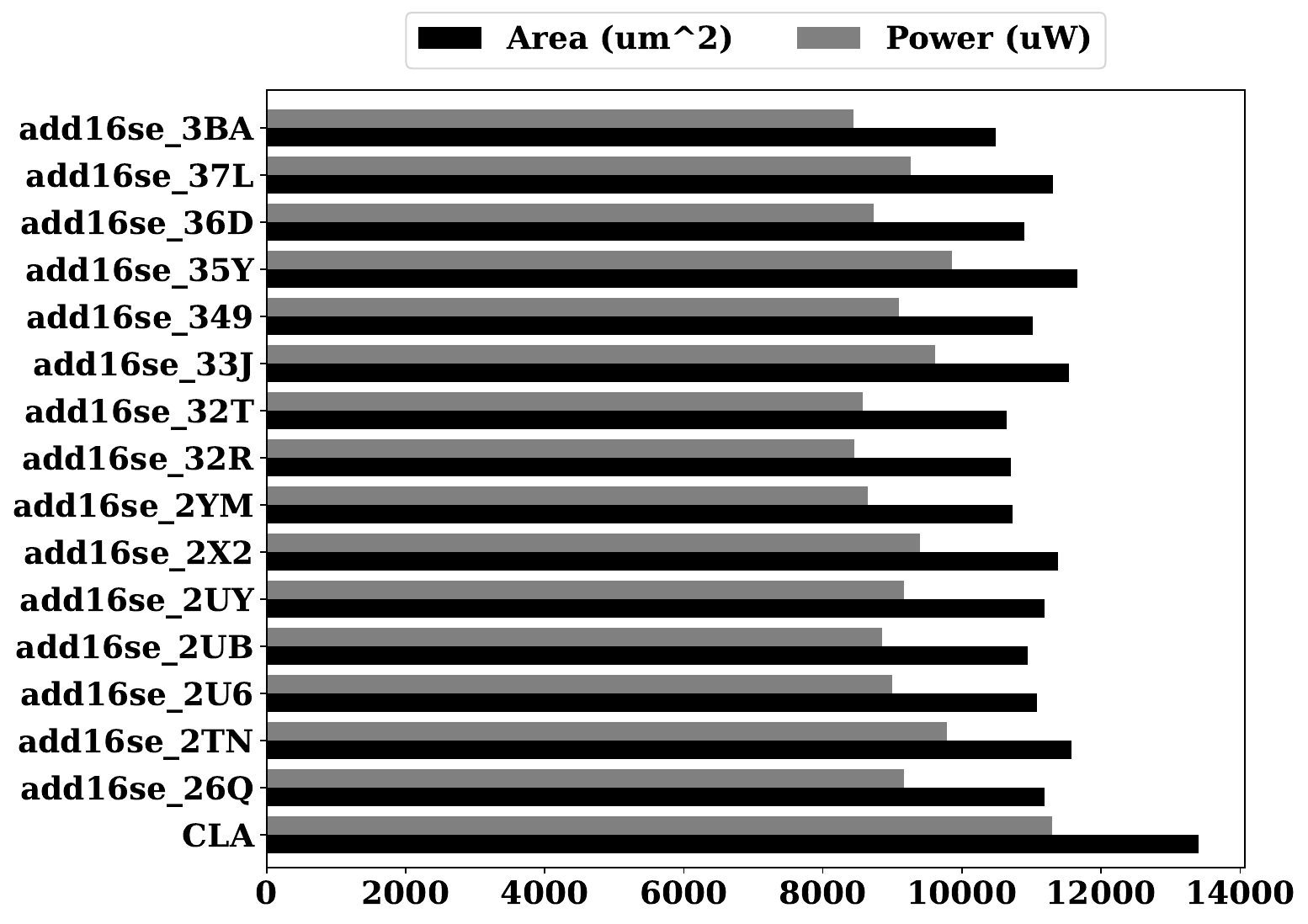}
    \caption{Area and power statistics for CORDIC core using both accurate and approximate adders.}
    \label{fig:hw}
    \vspace{-5ex}
\end{figure}
\subsection{Hardware Evaluation}
Our hardware evaluation uses Verilog HDL to describe a CORDIC core with approximate addition operations. 
We synthesize the core using  Synopsys Design Compiler with the NanGate $45nm$ Open Cell Library to obtain area and power statistics. 
The core runs at 500 MHz clock frequency.
The Carry-Lookahead Adder (CLA) is selected as the baseline accurate adder to match the high frequency requirement of 500 MHz. 

Figure~\ref{fig:hw} shows the area and power statistics obtained for the CORDIC core using both accurate and approximate adders. 
We make 2 major observations:
(1) All approximate adders perform better than the baseline (CLA) adder in consideration. 
(2) The approximate adder \texttt{add16se\_3BA} provides the best area and power savings, saving 21.74\% area and 25\% power compared to CLA. 
On average, we see that all the approximate adders 
%in consideration 
save 17.25\% on-chip area and 19.4\% power when compared to CLA. 
\vspace{-2ex}
    
\begin{figure}
    \includegraphics[width=0.5\textwidth]{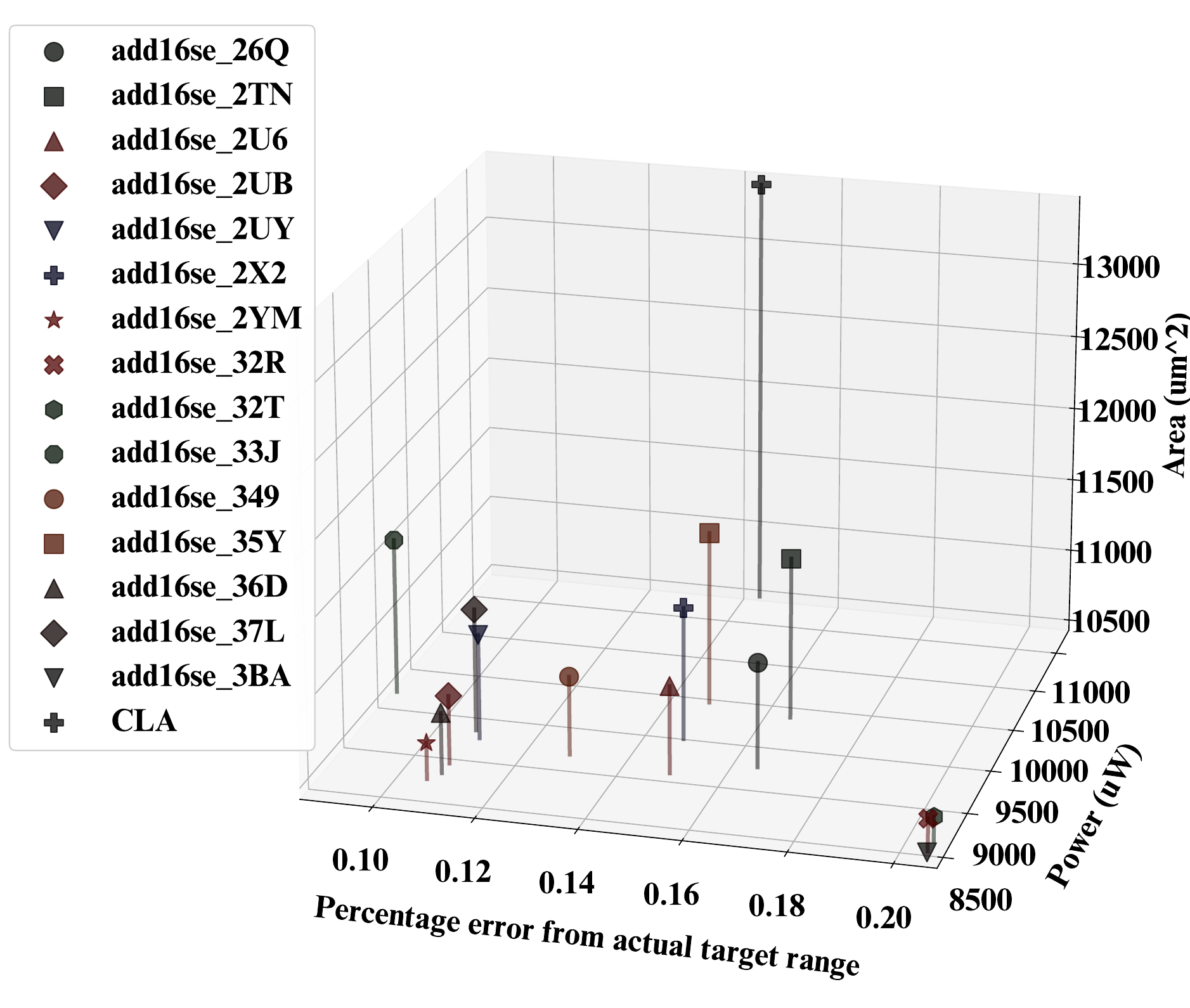}
    \caption{Design Space Exploration. Relationship between accuracy, area and power while using accurate and approximate adders in the presented scenario.}
    \label{fig:dse_nlp}
    \vspace{-3ex}
\end{figure}
\subsection{Design Space Exploration}
To study the relationship between accuracy and hardware statistics, \papername{} constructs a design space and explores optimal design points, i.e., adders with diverse accuracy and hardware metrics applied to this scenario so as to satisfy user-defined quality constraints. 
Figure~\ref{fig:dse_nlp} shows the design space of relationships between accuracy, area and power when different adders are applied to our end-to-end OFDM radar scenario. % presented in this paper. 
The accuracy metric is the average percentage error from the target's actual distance (50 m) in the positive SNR range (5 to 15 dB). 
The area and power metrics are the same as reported in Figure~\ref{fig:hw}. 
The design space in Figure~\ref{fig:dse_nlp}
can be used to evaluate different OFDM radar use cases.
%construct various cases. 
On one hand, if the user sets a quality constraint where maximum percentage error is 0.1\% in the positive SNR region, then from the DSE we see that adder \texttt{add16se\_33J} fulfills the criteria with a percentage error of 0.088\%. 
However, we observe a trade-off in area and power needs, since the area and power savings are 13.9\% and 14.6\% respectively which are low compared to other adders. 
On the other hand if the user sets a multi-objective optimization of a percentage error below 0.12\%, and an area and power savings above 18\% and 21\% respectively, then we have the adders \texttt{add16se\_2UB}, \texttt{add16se\_2YM}, \texttt{add16se\_36D} fulfilling those criteria. 

Therefore, based on accuracy requirements, coupled with an area and power budget, many observations can be made and \papername{} enables exploration of such optimal design points.

\section{Conclusion and Future Work}
We presented {\papername}, an approach exploiting approximations in MUSIC to enable design space exploration of alternative  accuracy-area-power tradeoffs for end-to-end OFDM radar pipelines.
The core contribution of {\papername} is the 
%. {\papername} investigates the use of approximate adders 
 use of approximate adders for the computationally intensive SVD block (CORDIC core) in MUSIC for an end-to-end OFDM radar pipeline. 
 %Particularly, {\papername} incorporates approximate adders inside the CORDIC algorithm which is used to compute SVD on hardware. 
 Our case study for an end-to-end OFDM radar pipeline demonstrated that 
 {\papername} can uncover an interesting design space
 that system designers can use to explore trade-offs and design alternatives.
 For instance, our experiments show that {\papername} helps save 17.25\% on-chip area and 19.4\% power with a minimal percentage error of 0.14\% on average in positive SNR.
Our ongoing and future work  considers integrating other algorithms such as~\cite{effmu1,effmu2} with \papername{} across an end-to-end pipeline to evaluate the combined effect on accuracy and hardware efficiency.
\bibliographystyle{IEEEtran}
\bibliography{References/sample}

\begin{thebibliography}{10}
\providecommand{\url}[1]{#1}
\csname url@samestyle\endcsname
\providecommand{\newblock}{\relax}
\providecommand{\bibinfo}[2]{#2}
\providecommand{\BIBentrySTDinterwordspacing}{\spaceskip=0pt\relax}
\providecommand{\BIBentryALTinterwordstretchfactor}{4}
\providecommand{\BIBentryALTinterwordspacing}{\spaceskip=\fontdimen2\font plus
\BIBentryALTinterwordstretchfactor\fontdimen3\font minus \fontdimen4\font\relax}
\providecommand{\BIBforeignlanguage}[2]{{%
\expandafter\ifx\csname l@#1\endcsname\relax
\typeout{** WARNING: IEEEtran.bst: No hyphenation pattern has been}%
\typeout{** loaded for the language `#1'. Using the pattern for}%
\typeout{** the default language instead.}%
\else
\language=\csname l@#1\endcsname
\fi
#2}}
\providecommand{\BIBdecl}{\relax}
\BIBdecl

\bibitem{music}
R.~Schmidt, ``Multiple emitter location and signal parameter estimation,'' \emph{IEEE Transactions on Antennas and Propagation}, vol.~34, no.~3, 1986.

\bibitem{use3}
A.~Fascista, G.~Ciccarese, A.~Coluccia, and G.~Ricci, ``A localization algorithm based on v2i communications and aoa estimation,'' \emph{IEEE Signal Processing Letters}, vol.~24, no.~1, pp. 126--130, 2017.

\bibitem{biomed}
Y.~Zhong, J.~Xiang, X.~Chen, Y.~Jiang, and J.~Pang, ``Multiple signal classification-based impact localization in composite structures using optimized ensemble empirical mode decomposition,'' \emph{Applied Sciences}, vol.~8, no.~9, 2018.

\bibitem{use2}
Y.~Zhou, J.~Li, H.~Yan, and X.~Yan, ``Low-frequency ultrasound thoracic signal processing based on music algorithm and emd wavelet thresholding,'' \emph{IEEE Access}, vol.~11, pp. 73\,912--73\,921, 2023.

\bibitem{seismic}
M.~Heck, M.~Hobiger, A.~van Herwijnen, J.~Schweizer, and D.~Fäh, ``{Localization of seismic events produced by avalanches using multiple signal classification},'' \emph{Geophysical Journal International}, vol. 216, no.~1, pp. 201--217, 09 2018.

\bibitem{musicuse}
X.~Chen, Z.~Feng, Z.~Wei, X.~Yuan, P.~Zhang, J.~A. Zhang, and H.~Yang, ``Multiple signal classification based joint communication and sensing system,'' \emph{IEEE Transactions on Wireless Communications}, vol.~22, no.~10, pp. 6504--6517, 2023.

\bibitem{pak}
U.~M. Butt, S.~A. Khan, A.~Ullah, A.~Khaliq, P.~Reviriego, and A.~Zahir, ``Towards low latency and resource-efficient fpga implementations of the music algorithm for direction of arrival estimation,'' \emph{IEEE Transactions on Circuits and Systems I: Regular Papers}, 2021.

\bibitem{admm}
A.~S. Assoa, A.~Bhat, S.~Ryu, and A.~Raychowdhury, ``A scalable platform for single-snapshot direction of arrival (doa) estimation in massive mimo systems,'' in \emph{Proceedings of the Great Lakes Symposium on VLSI 2023}, 2023, pp. 631--637.

\bibitem{snap}
G.~A. Ioannopoulos, D.~E. Anagnostou, and M.~T. Chryssomallis, ``Evaluating the effect of small number of snapshots and signal-to-noise-ratio on the efficiency of music estimations,'' \emph{IET Microwaves, Antennas \& Propagation}, vol.~11, no.~5, pp. 755--761, 2017.

\bibitem{onebit}
X.~Huang and B.~Liao, ``One-bit music,'' \emph{IEEE Signal Processing Letters}, vol.~26, no.~7, pp. 961--965, 2019.

\bibitem{surveyapprox}
S.~Mittal, ``A survey of techniques for approximate computing,'' \emph{ACM Comput. Surv.}, vol.~48, no.~4, mar 2016.

\bibitem{cesa}
\BIBentryALTinterwordspacing
R.~Bhattacharjya, V.~Mishra, S.~Singh, K.~Goswami, and D.~S. Banerjee, ``An approximate carry estimating simultaneous adder with rectification,'' in \emph{Proceedings of the 2020 on Great Lakes Symposium on VLSI}, ser. GLSVLSI '20.\hskip 1em plus 0.5em minus 0.4em\relax New York, NY, USA: Association for Computing Machinery, 2020, p. 139–144. [Online]. Available: \url{https://doi.org/10.1145/3386263.3406928}
\BIBentrySTDinterwordspacing

\bibitem{locate}
\BIBentryALTinterwordspacing
R.~Bhattacharjya, B.~Maity, and N.~Dutt, ``Locate: Low-power viterbi decoder exploration using approximate adders,'' in \emph{Proceedings of the Great Lakes Symposium on VLSI 2023}, ser. GLSVLSI '23.\hskip 1em plus 0.5em minus 0.4em\relax New York, NY, USA: Association for Computing Machinery, 2023, p. 409–413. [Online]. Available: \url{https://doi.org/10.1145/3583781.3590314}
\BIBentrySTDinterwordspacing

\bibitem{svdresil}
P.~C. Hansen, ``The truncated svd as a method for regularization,'' \emph{BIT Numerical Mathematics}, vol.~27, pp. 534--553, 1987.

\bibitem{cordic}
R.~Andraka, ``A survey of cordic algorithms for fpga based computers,'' in \emph{Proceedings of the 1998 ACM/SIGDA Sixth International Symposium on Field Programmable Gate Arrays}, ser. FPGA '98, 1998.

\bibitem{ref_golub}
G.~Golub and W.~Kahan, ``Calculating the singular values and pseudo-inverse of a matrix,'' \emph{Journal of the Society for Industrial and Applied Mathematics Series B Numerical Analysis}, vol.~2, no.~2, 1965.

\bibitem{evoapprox}
V.~Mrazek, L.~Sekanina, and Z.~Vasicek, ``Libraries of approximate circuits: Automated design and application in cnn accelerators,'' \emph{IEEE Journal on Emerging and Selected Topics in Circuits and Systems}, vol.~10, no.~4, pp. 406--418, 2020.

\bibitem{stat}
Z.~Vasicek, ``Formal methods for exact analysis of approximate circuits,'' \emph{IEEE Access}, vol.~7, pp. 177\,309--177\,331, 2019.

\bibitem{effmu1}
K.~Todros and A.~O. Hero, ``Robust multiple signal classification via probability measure transformation,'' \emph{IEEE Transactions on Signal Processing}, vol.~63, no.~5, pp. 1156--1170, 2015.

\bibitem{effmu2}
F.~Chen, D.~Yang, and S.~Mo, ``A doa estimation algorithm based on eigenvalues ranking problem,'' \emph{IEEE Transactions on Instrumentation and Measurement}, vol.~72, pp. 1--15, 2023.

\end{thebibliography}

\end{document}